\documentstyle[prd,aps]{revtex}

\newcommand{\be}{\begin{equation}}
 \newcommand{\ee}{\end{equation}}
 \newcommand{\beqa}{\begin{eqnarray}}
 \newcommand{\eeqa}{\end{eqnarray}}

\begin{document}
\draft

\title{A possible Newtonian interpretation of relativistic cosmological perturbation theory}
\author{{Ali ~Nayeri\thanks{ali@iucaa.ernet.in},$\;$ and$\;$ 
T.~Padmanabhan\thanks{paddy@iucaa.ernet.in}}\\
{{\it IUCAA, Post Bag 4, Ganeshkhind, Pune 411 007, INDIA.}}}

\maketitle
\begin{abstract}
\parshape=1 0.75in 5.5in
Cosmological perturbations with wavelengths smaller than Hubble radius can be 
handled in the context of Newtonian theory with very high accuracy. The application of this Newtonian approximation, however, is restricted to nonrelativistic matter and cannot be used for relativistic 
matter. Recently,
by modifying the continuity equation, Lima, et. al., extended the domain
of applicability of Newtonian cosmology to radiation dominated phase.
We adopted this continuity equation to re-examine
linear cosmological perturbation theory for a two fluid universe with uniform pressure. We study the  evolution equations 
for density contrasts and their validity in different epochs and on scales larger than Hubble radius and compare the results with the full relativistic approach.
The comparison shows the high accuracy of this approximation.
\end{abstract}
\section{introduction}
The real universe contains inhomogeneous structures like galaxies,
clusters etc and, in any theory of the formation of these structures, it is 
essential to understand the evolution of small inhomogeneities in the early
universe. In principle, it is straightforward to work out the general relativistic theory of linear
perturbations~\cite{lif1,lif2}. By linearizing Einstein's equations,  we can
 obtain a second-order differential equation of the form
\be
\hat{\cal L}(g_{\alpha\beta}) \ \delta g_{\alpha\beta} = \delta T_{\alpha\beta},
\ee
where $\hat{\cal L}$ is a linear differential operator depending on the
background metric, $g_{\alpha\beta}$, and the set $(\delta g_{\alpha\beta} , 
\delta T_{\alpha\beta})$ denotes the perturbations in the metric and stress
tensor about an expanding FRW background.

In practice,  however,  there are many complications and coneptual difficulties
which make this analysis highly nontrivial. One issue is the so-called
``gauge problem'' ~(extensively discussed in the literature; see [3-13]) which arises 
due to nonuniqueness of splitting the metric and matter variables into a zeroth 
order background and small, first-order perturbations. Since a
relabeling of coordinates $x^{\alpha} \rightarrow x^{\alpha'}$ can
make a small $\delta T_{\alpha\beta}$ large or even generate a component
which was originally absent,  
one must care to factor out effects due to coordinate transformations,  when analyzing 
relativistic perturbations. There
are two different ways of handling these difficulties in general 
relativity. One approach is to analyze a perturbation in a particular gauge,
say, synchronous gauge~\cite{wein}. In this case, one specifically identifies
the points of a fictitious background spacetime with those of real spacetime,
and treat $\delta{T^0}_0$ to be the perturbed mass density etc.
In this method, however, we cannot fix the gauge completely and the
residual gauge ambiguities can create some problems. The second method is to
construct the perturbed physical variables in a gauge-invariant manner.
The gauge-invariant approach is conceptually more attractive since there is no need 
for specific identification of the points between the two spacetimes, though it is 
more complicated and the physical meaning of variables  do not, in general,
posses any simple interpretation and becomes obvious only for specific
observers.

It is convenient to divide the cosmological perturbations into two subclasses:
(i) Perturbations with wavelengths larger than Hubble scale ($\lambda > d_H$), for 
which we have to use some form of a
general relativistic theory of perturbations and (ii) small-scale perturbations 
($\lambda < d_H$) for which the evolution of mass density can be studied
using Newtonian theory. In this context, all physical quantities can be defined 
unambiguously to the order of accuracy needed. In general, the application of Newtonian equations
is further restricted to nonrelativistic matter and cannot be used for relativistic 
component even for scales
much smaller than Hubble radius ($\lambda\ll d_H)$.

Recently, Lima~\cite{lima}, et al., re-examined the basic equations describing
a Newtonian universe with {\it uniform pressure} and found a way for obtaining the 
same evolution equation for density contrast as could be obtained by the full relativistic 
approach. They achieved  this goal by modifying the
continuity equation in an expanding background. 
Using this result,
 they argued that one can extend the domain of validity of
Newtonian cosmology~\cite{mm} in order to analyze some problems of formation of
structures even in the radiation dominated phase.

In this paper we extend the result of reference~\cite{lima} to a multi-component
universe with different equations of state. We shall then consider a
two fluid universe in the context of ``{\it pseudo Newtonian}''cosmology.
Comparison with the fully relativistic two-component universe reveals
the high accuracy of density contrast equations in the pseudo Newtonian
cosmology.

The organization of this paper is as follows: In section II we will
use the modified continuity equation in order to find out the evolution
equation for density contrast of a multi-component universe with arbitrary
equations of state. This is actually the generalization of paper by
Lima~\cite{lima} , et. al. In section III we will examine the result of
section II for the two fluid system with radiation and dark
matter. In this section we will try to give 
approximate solutions to coupled equations of density contrast of
radiation and dark matter. We shall compare the result in this
context with those ones as obtained in fully relativistic approach,
for example given in ref~\cite{paddy1}. We give the conclusions in section IV.

\section{The linear perturbation theory for a multi-component universe}
The set of equations describing a several component universe, in the proper 
coordinates $(t, {\bf r})$ with ${\bf r} = a(t){\bf x}$, can be written as,
 (i) {\it ``Continuity''} equation, (ii){\it Euler} equation, and (iii) {\it Poisson}
 equation,
\be
\frac{\partial \rho}{\partial t}+\nabla_r\cdot(\rho {\bf u})
+p\nabla_r\cdot{\bf u}  =  0, \label{nc}
\ee
\be
\frac{\partial {\bf u}}{\partial t}+({\bf u}\cdot\nabla_r){\bf u}
 =  -\nabla_r\psi-(\rho+p)^{-1}\nabla_r p,   
\ee
\be
{\nabla_r}^2\psi  =  4\pi G(\rho+3p), 
\ee
where the ``{\it continuity}'' equation is the one that has been used
by Lima~\cite{lima}, et. al., for an expanding universe, to
study the evolution of cosmological perturbation using linear regime. Here,
the velocity of light, $c$, assumed to be unity and the quantities
\be
\rho = \sum_{i=1}^N \rho_i,\hspace*{0.5cm} p = \sum_{i=1}^N  p_i, \hspace*{0.5cm} {\bf u} = \sum_{i=1}^N
{\bf u_i} \hspace*{0.5cm} and \hspace*{0.5cm}
\psi = \sum_{i=1}^N  \psi_i
\ee
denote the total density, total pressure, total
field velocity, and the total generalized gravitational potential of the
cosmic fluid. 

In an expanding Newtonian universe the evolution of small fluctuations
can be studied in the usual perturbation theory, i.e, by perturbing
the background density, pressure, field velocity, and gravitational 
potential. That is, we take
\beqa
&&\rho ({\bf r}, t) = \rho_{bg}(t)[1+\delta ({\bf r}, t)],  \\
&&p ({\bf r}, t) = p_{bg}(t)+\delta p({\bf r}, t),  \\
&&\psi ({\bf r}, t) = \psi_{bg}({\bf r}, t)+\varphi ({\bf r}, t),  \\
&&{\bf u}({\bf r}, t) = {\bf u}_{bg}+{\bf v}({\bf r}, t),
\eeqa

where the second term is a small correction. By inserting the above expressions 
into Eqs (1), linearizing
the resulting equations to first order in perturbations, and
transforming to comoving coordinates by ${\bf x} = ({\bf r}/a)$,
we get,
\beqa
&&\dot\delta^{(X)}+\frac{1+\nu^{(X)}}{a}\nabla_x\cdot{\bf v}^{(X)}
 = 0 \nonumber \\
&&\dot{\bf v}^{(X)}+\frac{\dot a}{a}{\bf v}^{(X)} = -\frac{1}{a}\nabla_x\varphi
-\frac{{{v_s}^2}^{(X)}}{(1+\nu^{(X)})a}\nabla_x\delta^{(X)} \\
&&{\nabla_x}^2\varphi = 4\pi G a^2 \sum_{i}(1+3\nu_i){\rho_{bg}}_i\delta_i \nonumber
\eeqa
where we have assumed that the expansion is adiabatic and thus 
${v_s^{(X)}}^2=(\delta p/\delta \rho) = \nu^{(X)}$, is the sound velocity. 
Here ($X$) denotes the
species under consideration and $p_i = \nu_i\rho_i$. The sum on the right hand side 
is over all components. Note that $\nu_{tot}$ and ${v_s}_{tot}^2$ are not
constant any more and thus the relation ${v_s}_{tot}^2 = \nu_{tot}$ does
not hold for the full system.

By eliminating the peculiar velocity between the above equations,
and Fourier transforming the perturbation such that 
${\nabla_x}^2\delta^{(X)} = -k^2\delta^{(X)}$ etc.,  we get
\be
{\ddot\delta_k}^{(X)}+2\frac{\dot a}{a}{\dot\delta_k}^{(X)}
+\left(\frac{k^{(X)}{v_{s}}^{(X)}}{a}\right)^2{\delta_k}^{(X)} =
 4\pi G (1+\nu^{(X)})\sum_i (1+3\nu_i){\rho_{bg}}_i{\delta_k}_i. \label{main} 
\ee

This is a peculiar equation in the sense that there is no counterpart for this equation 
in the fully relativistic treatment, unless we impose the synchronous gauge and comoving 
gauge simultaneously. (one cannot impose the two gauge conditions simultaneously even in 
large scale limit in the presence of pressure, since we get $\delta = 0$ system when we 
ignore the entropic and anisotropic pressures~\cite{jhwang}.) In the case of single component 
medium~\cite{lima} and $(k^2\ {v_s}^2/a^2) \ll 1$,  however,  equation~(\ref{main}) 
leads to the same equation derived in full relativistic approach by imposing
the two gauge conditions simultaneously (see for
example, eq. (15.10.57) in ref.~\cite{wein}, eq. (10.118) in ref.~\cite{peebles} and 
eq. (6.136) in ref.~\cite{paddy2}).

\section{two fluid universe}
Now, we shall assume that our universe consists of radiation [R] and
dark matter [DM] with  $\nu_R = 1/3 = v_R^2$ and $\nu_{DM} = v_{DM}^2$ = 0. 
Then the evolution of $\delta^R_k(t)$ and $\delta^{DM}_k(t)$ can be determined 
from eq.~(\ref{main}), to give
\beqa
&&\ddot\delta_R+2\frac{\dot a}{a}\dot\delta_R+\left[\frac{k^2v_R^2}{a^2}
-\frac{32\pi G}{3}\rho_R\left(1+\frac{\delta\rho_{DM}}{2\delta\rho_R}
\right)\right]\delta_R=0. \\
&&\ddot\delta_{DM}+2\frac{\dot a}{a}\dot\delta_{DM}
-4\pi G\rho_{DM}\left(1+\frac{2\delta\rho_R}
{\delta\rho_{DM}}\right)\delta_{DM}=0. 
\eeqa
Though these equations cannot be solved exactly in closed form
all the important properties of the solutions can be obtained by
suitable approximations. We shall now discuss these properties
by introducing the new variable $x \equiv (a/a_{eq})$ where $a_{eq}$ is the
expansion factor at the time $ t = t_{eq}$. Transforming from $t$ to $x$ as
independent variable we can re-express all the quantities in terms of x,
\be
\frac{\rho_{DM}}{\rho_{eq}}= \frac{1}{2x^3}; \hspace{0.5cm}
\frac{\rho_R}{\rho_{eq}}= \frac{1}{2x^4}; \hspace{0.5cm}
\frac{\rho_{tot}}{\rho_{eq}}= \frac{1}{2x^4}(x+1).
\ee
 Since $(p_{tot}/{\rho_{eq}}) = (p_{DM}+p_R)/{\rho_{eq}}=(1/6x^4)$
we get the following result for the full two component system
\be
\nu= \nu_{tot}= \frac{p_{tot}}{\rho_{tot}} = \frac{p_{DM}+p_R}{\rho_{DM}+
\rho_R} = \frac{1}{3 (1+x)},
\ee
and
\be
H^2(x) = H^2_{eq} \frac{1}{2x^4}(1+x).
\ee 
On the other hand, one can express the combination $(k^2/H^2a^2)$ as
\be
\frac{k^2}{H^2a^2} = \frac{2x^2}{(1+x)}\omega^2,
\ee
where $\omega $ is defined by the ratio $2\pi\omega = [d_H(a_{eq})/\lambda
(a_{eq})]$.  In terms of this variable, $x$, the time derivatives can be given by
\beqa
\frac{d}{dt} & = & Ha\frac{d}{da} = H(x) x \frac{d}{dx} \equiv H \hat D \nonumber \\
\frac{d^2}{dt^2} & = & H^2 a \frac{d}{da}\left(a\frac{d}{da}\right)
-\frac{3}{2}H^2(1+\nu)a \frac{d}{da}\equiv H^2\hat D^2-\frac{3}{2}H^2(1+\nu)
\hat D,
\eeqa
where $\hat D = x(d/dx)$ and we have used the new continuity equation~(\ref{nc})
. With these modifications, the coupled equations for $\delta_R$ and
$\delta_{DM}$ become
\be
\left[\hat D^2+\frac{1}{2}\frac{x}{(1+x)}\hat D+\left(\frac{2}{3}\frac{\omega^2x^2}
{(1+x)}-\frac{4}{(1+x)}\right)\right] \delta_R = \frac{2x}{(1+x)}\delta_{DM},
\label{r}
\ee
and
\be
\left[\hat D^2+\frac{1}{2}\frac{x}{(1+x)}\hat D-\frac{3}{2}\frac{x}{(1+x)}\right]\delta_{DM}=
\frac{3}{(1+x)}\delta_R. \label{dm}
\ee

A particular mode, labeled by the parameter $\omega$, enters the Hubble
radius in the radiation dominated phase if $\omega>1$ and in matter dominated
phase if $\omega<1$. The $\omega$ and $x_{ent}$ are related to each other through 
the relation $\omega^2x_{ent}^2 = 2\pi^2(1+x_{ent})$. We shall
consider the case $\omega^2\gg 1$ which has more features and is more
relevant to this paper. Then three cases need to be discussed.
(i) When $x\omega\ll 1$, in which case the wave length of the mode is bigger than 
the Hubble radius;
(ii) when $\omega^{-1}\ll x\ll 1$, the mode has entered the Hubble radius in the
radiation dominated phase; and finally the case, (iii) when $\omega x \gg 1$, the 
mode is inside the Hubble radius in a matter dominated phase.  Let us now
consider each case.

\centerline{{\bf \underline{Case (i) $x \omega \ll 1$ }}}
 
In this case the equation~(\ref{r}) and~(\ref{dm}) 
give
\be
\left(\hat D^2 - 4\right)\delta_R \cong 0; \hspace{0.5cm} \hat D^2\delta_{DM}\cong 3 \delta_R.
\ee
The growing solution is
\be
\frac{3}{4}\delta_R = \delta_{DM} = x^2
\ee
Thus both radiation and dark matter grow as $x^2 \propto a^2$. Obviously, since
the universe is radiation dominated $\delta_{DM}$ is driven by $\delta_R$ and
its back reaction on $\delta_R$ is negligible. This result matches
with the fully relativistic one!

\centerline{{\bf \underline{Case (ii) $\omega x \gg 1$ and $x\ll 1$}}} 

The concerned equations in this approximation are
\be
\left[\hat D^2+\frac{2}{3}\omega^2x^2\right]\delta_R\cong0;\hspace{0.5cm}
\hat D^2\delta_{DM}\cong3\delta_R
\ee
The solution for $\delta_R$ equation would be of the form
\be
\delta_R = C_1 \  J_0\left[ \sqrt{\frac{2}{3}} \  \omega x\right]+
C_2 \ N_0\left[ \sqrt{\frac{2}{3}}\ \omega x\right],
\ee
where $J_0$ and $N_0$ are the Bessel functions of the first and second kind, respectively. 
If we use the fact that $\omega x\gg
1$ the approximate solution to $\delta_R$ becomes
\be
\delta_R \simeq \frac{C}{\sqrt{\omega x}} \sin{\left(\sqrt{\frac{2}{3}} \ \omega x\right)},
\ee
which is oscillating rapidly since $\omega x \gg 1$. Substituting for $\delta_R$
in the second equation and using the fact that $x \ll 1$, we find that
\be
\delta_{DM} \simeq  -\frac{6 C}{\sqrt{\omega x}} \sin{\left(
\sqrt{\frac{2}{3}} \ \omega x\right)} + A \ln{x} + B.
\ee
Therefore the perturbation in dark matter grows essentially logarithmically during 
the period $\omega^{-1} \ll x \ll 1$, which also known from previous work. 
  Though the wavelength of the dark matter perturbation is bigger than effective 
Jeans length, the growth of dark matter perturbation is prevented by
the rapid expansion of the universe. The similarities of these results
with their fully relativistic ones come from using the modified continuity
equation,  which-in turn-permits one to enlarge the domain of applicability of 
Newtonian cosmology even to radiation dominated phase.

\newpage
\centerline{{\bf \underline{Case (iii) $\omega x \gg 1$}}}

 This range corresponds to the matter dominated
phase with mode inside the Hubble radius. The coupled equations now become
\be
\left[\hat D^2+\frac{1}{2}\hat D-\frac{3}{2}\right]\delta_{DM}\cong 0; \hspace{0.5cm}
\left[\hat D^2+\frac{1}{2}\hat D+\frac{2}{3}\omega^2x\right]\delta_R\cong 2\delta_{DM}
\ee
Solving for $\delta_{DM}$ we get
\be
\delta_{DM} = Ax+Bx^{-2/3} \cong Ax.
\ee
Plugging this into the $\delta_R$ equation we find that
\be
\frac{d}{dx}\left(x\frac{d\delta_R}{\delta x}\right)+\frac{1}{2}
\frac{d\delta_R}{dx}+\frac{2}{3}\omega^2\left(\delta_R-\frac{3A}{\omega^2}
\right)=0
\ee
Choosing, $\delta_R=(3A/\omega^2)+x^{-3/4}f(x)$, we get the following
differential equation for $f(x)$
\be
f''(x)= -\left(\frac{3}{16}\frac{1}{x^2}+\frac{2}{3}\frac{\omega^2}{x^2}\right)f
(x)\cong -\frac{2}{3} \ \frac{\omega^2}{x}f(x).
\ee
By applying the WKB approximation, we may solve the $f$ equation and get
the final result for $\delta_R$ to be
\be
\delta_R = \frac{3A}{\omega^2}+B\frac{1}{\sqrt{\omega x}} \ \exp\left(
\pm i \sqrt{\frac{8}{3}} \ \omega x^{1/2}\right).
\ee
The oscillations presented by the second term, in $\delta_R$, continue to dominate over 
the driving by the $\delta_{DM}$
term. Hence, the radiation density does not grow in the matter dominated universe, i.e., 
when $\lambda_R < d_H$. These solutions are very similar to the solutions to the fully 
relativistic equations. The
reasons are clear during this phase: First of all, we are dealing with the modes which are 
well within the Hubble radius and secondly the modes are in the matter dominated phase.
\section{conclusions}
We have shown that the result of the linear cosmological perturbation of a two fluid 
universe can be obtained with high accuracy from Newtonian cosmology even in the presence 
of pressure and for scales larger than Hubble radius. This is obtained by using the modified 
Newtonian equations. In fact, by using the modified continuity equation one can get rid of a 
misleading pressure gradient term which is obtained in the semi-classical formulation and more
 over obtain the time evolution of the density contrast for any value of parameter $\nu$.   
Comparison of~(\ref{r}) and~(\ref{dm}) with fully relativistic equations, say, in the comoving 
gauge (see e.g. ref.~\cite{paddy1}) given by:
\be
\left[\hat D^2+\frac{1}{2}\frac{x}{(1+x)}\hat D+\left(\frac{2}{3}\frac{\omega^2x^2}
{(1+x)}-\frac{4}{(1+x)}\right)\right] \delta_R - \frac{2x}{(1+x)}\delta_{DM}
= F(\delta_R, \delta_{DM}) ,\label{r21}
\ee
and
\be
\left[\hat D^2+\frac{1}{2}\frac{x}{(1+x)}\hat D-\frac{3}{2}\frac{x}{1+x}\right]\delta_{DM} -
\frac{3}{1+x} \delta_R = G(\delta_R)\label{r22},
\ee
with
\be
F(\delta_R, \delta_{DM}) = \left[\hat D - \frac{4}{9}\frac{6 x^2 + 13 x +8}
{(x+4/3)^2 (1+x)}\right] \delta_R - \frac{4}{3}\frac{x}{(x+4/3)}\hat D \delta_{DM}, \label{F}
\ee
and
\be
G(\delta_R) = \frac{1}{x+4/3}\left[\hat D - \frac{1}{3}\frac{6 x^2 + 13 x + 8}
{(1+x) (x+4/3)}\right] \delta_R, \label{G}
\ee
will allow us to estimate the accuracy. Left hand sides of Eqs.~(\ref{r21}) and~(\ref{r22}) are
the same as their pseudo Newtonian counterparts Eqs.~(\ref{r}) and~(\ref{dm}), while
$F$ and $G$ are the corrections coming from general relativity. Numerical comparison of $F$ and $G$ 
with their left hand sides in Eqs.~(\ref{r}) and
(\ref{dm}) reveals that these corrections are very small and almost negligible indeed. 
This can be better seen in Figs.~1 and 2. In each of these figures
the vertical axis is the ratio of pseudo Newtonian density contrast to
its exact relativistic value at the same epoch. Both the figures show that
this ratio for the radiation component is almost unity and the
value of dark matter density contrast in pseudo Newtonian case differs at
most by a factor $2$.  As we
discussed earlier, the solutions of Eqs.~(\ref{r}) and~(\ref{dm}) for three
different cases are in a very good agreement with the solutions of Eqs.~(\ref{r21}) 
and~(\ref{r22}). This agreement for $\omega \gg 1$ is more 
profound (see Fig.~1 and Fig.~2). In other words, the ratio $R = (\delta_{Newton}/\delta_{exact})$ for radiation tends to unity for all
epochs as we increase the value of $\omega$, though 
remains off for dark matter by a lesser factor.  It seems that the Friedmann's
equations have strong correspondence with  Newtonian theory, even more than one might have naively expected.

AN wishes to thank Robert Brandenberger for many useful discussions.

\newpage
\begin{center}
{\bf Figure captions}
\end{center}
\noindent {\bf Fig. 1.} Comparison of perturbation amplitudes in dark
matter (DM) in {\it pseudo Newtonian cosmology} (PNC) and {\it relativistic cosmology} (RC). The amplitude of DM in NC is more than RC with an error of less than
factor $2$ for $\omega = 100$. After ${\it z} = 390$ up to present time the
ratio becomes constant. The radiation amplitude in RC is closer to that in NC. The behavior of two
perturbations are almost the same except for some small phase difference.
The ratio $R = (\delta_{Newton}/\delta_{exact})$ for radiation is almost one
except for some tiny fluctuations.

\noindent {\bf Fig. 2} This is same as Fig. 1 with $\omega = 200$.
Note that the agreement between the modes are more pronounced. The ratio
for dark matter in this case is less than 2 and it becomes a constant
after ${\it z} = 39$ . In
radiation dominated phase when modes are inside the Hubble radius,
modes are almost in phase and again amplitude of radiation
is more in RC compare to PNC at any instant.

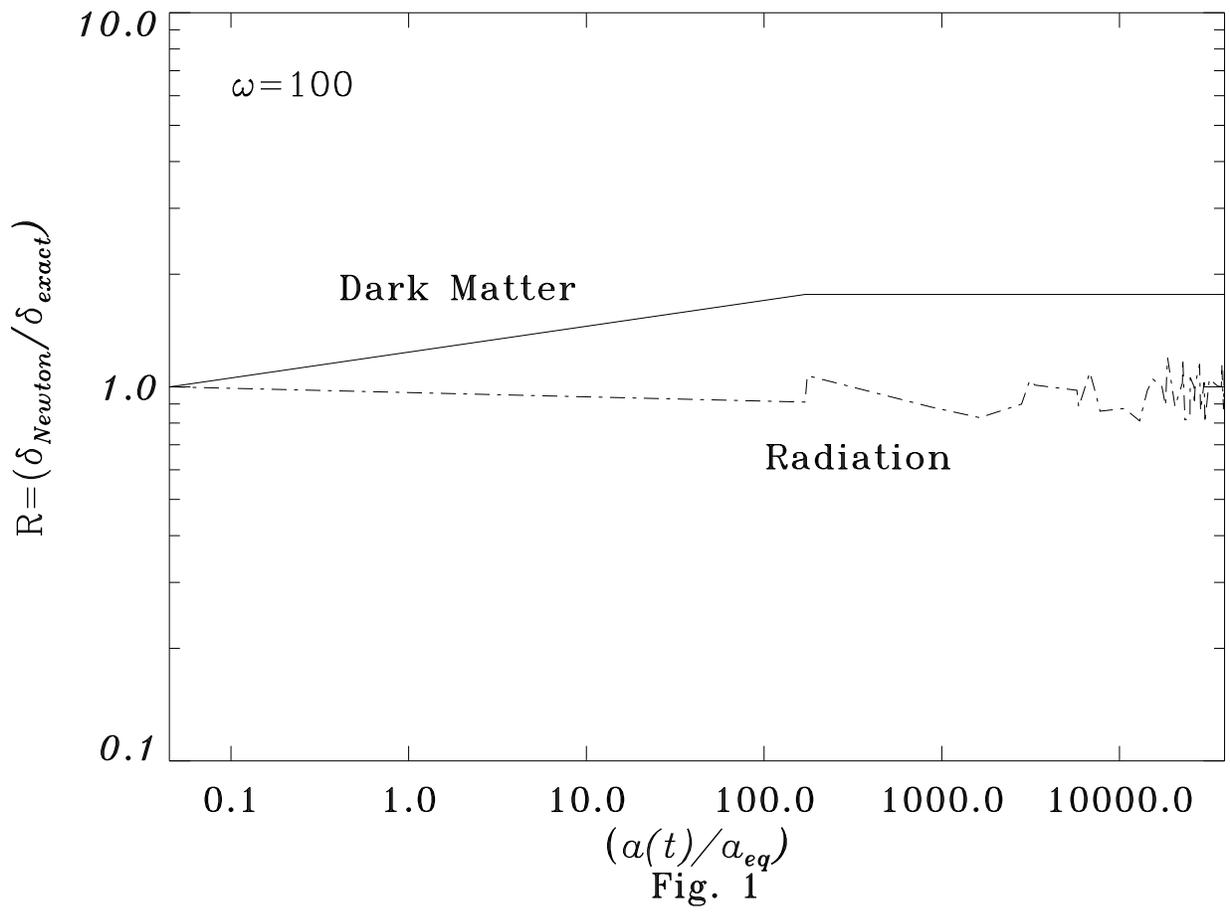
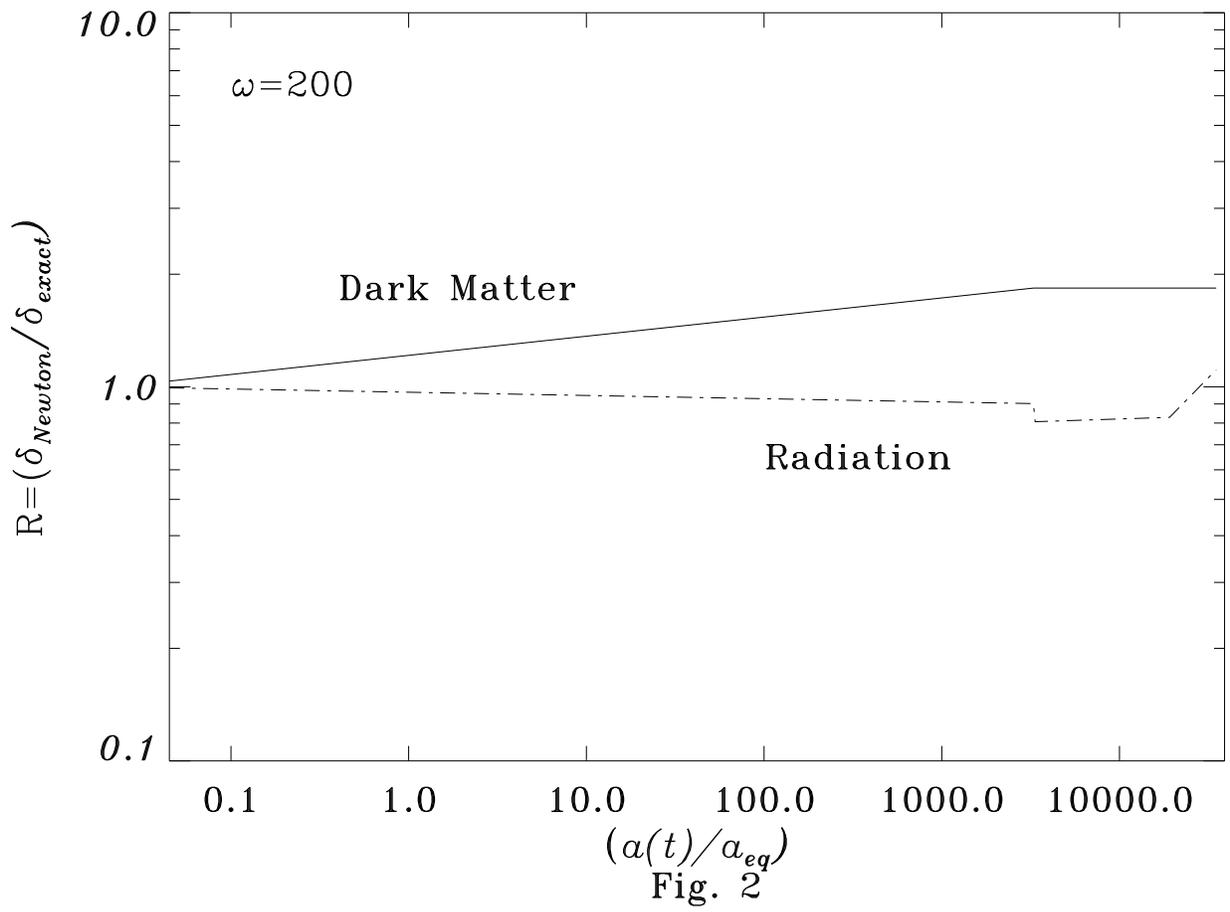
\end{document}